\documentclass[11pt]{elsarticle}

\usepackage[margin=1.25in]{geometry}

\usepackage{amsmath}
\usepackage{amssymb}
\usepackage{graphicx}
\usepackage{amsfonts}
\usepackage{dutchcal}
\usepackage{braket}
\usepackage{enumitem}

\usepackage{tikz}
\usepackage{forest}
\usetikzlibrary{trees}
\usetikzlibrary{calc}
\usepackage{calculator}
\usepackage{standalone}

\begin{document}

\title{Hamiltonian mechanics is conservation of information entropy}
\author{Gabriele Carcassi*, Christine A. Aidala \\ Physics Department, University of Michigan \\ 450 Church St, Ann Arbor, MI 48109, USA \\
	Corresponding author: Gabriele Carcassi, carcassi@umich.edu}

\begin{abstract}
	In this work we show the equivalence between Hamiltonian mechanics and conservation of information entropy. We will show that distributions with coordinate independent values for information entropy require that the manifold on which the distribution is defined is charted by conjugate pairs (i.e.~it is a symplectic manifold). We will also show that further requiring that the information entropy is conserved during the evolution yields Hamilton's equations.
\end{abstract}
\maketitle

\tableofcontents
\newpage

\section{Introduction}

Over the past decade there has been a renewed interest in the underpinning of classical mechanics. Different attempts and arguments have been put forward to understand what are the most important features of the various mathematical formulations, whether they are equivalent or whether one is more fundamental than the other~\cite{North,Curiel,Barrett1,Barrett2}. We believe, though, that there are two general problems with these attempts.

The first is that they concentrate on the mathematical structure which, unfortunately, is not enough to characterize a physical theory. By mathematical structure here we mean the part of the theory that is specified by formal relationships among abstract symbols, while by physical content we mean the link, necessarily not formal, between those symbols and actual physical concepts which, ultimately, must connect with experimental verification at least potentially. The problem is that the same physical content can be described by different mathematical frameworks. For example, as it is noted in~\cite{North,Curiel,Barrett1,Barrett2}, the state of many classical systems can be characterized by a point in a tangent bundle (through position and velocity) in the Lagrangian framework and by a point in a cotangent bundle (through position and momentum) in the Hamiltonian framework. Conversely, the same mathematical structure can be attached to different physical content. For example, the equation $ma + bv = 0$ in the context of Newtonian mechanics can either represent a body in an inertial frame under linear drag or a body in a non-inertial frame with no forces acting on it. If $t_{in}$ is time in the inertial frame, the time $t$ in the non-inertial frame is such that  $\frac{dt}{dt_{in}} = e^{\frac{b}{m}t}$. In other words, we can have the same physics with different math and the same math with different physics. The connection between math and physics is many-to-many because mathematics captures only the syntax of a particular description without its semantics, which at that point cannot be recovered. Therefore, if we want to analyze, compare or just understand a physical theory, it is not enough to understand the mathematical structure: we need a clear dictionary as to what each mathematical symbol ultimately represents in the physical model. That is, we should be able to ask simple questions such as: what does the numeric value of the Lagrangian for a given position and velocity represent? What does the numeric value of the symplectic form for two vectors represent?

If we just use category theory techniques, as in \cite{Barrett2}, to compare two physical theories, we cannot do anything else but compare the formal aspects. We may find that the formal parts of theories are inequivalent, but it could be that, once we add all the physical content that is needed to connect the formal part to actual physical systems and their measurement, they turn out to be equivalent; for similar arguments see~\cite{Coffey}. Or we may compare the formal structures of symplectic manifolds and Riemannian manifolds, and conclude, as in \cite{North}, that Hamiltonian mechanics is more fundamental than Lagrangian mechanics because defining only a volume element constrains the space less than defining only a line element. But, again, without having discussed exactly how that structure maps to the actual physical objects we cannot be sure of that conclusion. For example, kinetic momentum $mv$ is what is actually measured since conjugate momentum $p$ is not a gauge-invariant quantity. So it may be that, once we reconnect the formal theory to the actual physically defined quantities, we are forced to implicitly reintroduce the structure we appeared to be missing. 

Fixating on the mathematical structure without fully developing the link to the physical model it represents can also lead to confusion. For example, it is known~\cite{AllHamFreeParticle} that any Hamiltonian system is, at least locally, mathematically equivalent to a free particle. That is, we can always find a canonical transformation (i.e.~a transformation that does not change the equation of motion) such that any Hamiltonian system, in the new coordinates, has the motion of a free particle for a finite amount of time. As~\cite{North} notes, only frame-independent relationships should be taken as fundamental. Does that mean that the Hamiltonian itself is not fundamental because it is not invariant under canonical transformations? Does it mean that all Hamiltonian systems are the same system, since locally there is always a set of canonical coordinates in which $H=p^2/2m$? It is also known~\cite{AllSystemsAreHam} that for any first-order system of ordinary differential equations we can find an appropriate symplectic structure such that the system can be written as a Hamiltonian system. Does that mean that, in the end, all systems are physically equivalent, regardless of what they describe? The point is, we cannot know whether our findings about mathematical structures are physically significant if we do not have a precise link between the mathematical symbols and their physical meaning.

This leads us to the second problem we believe the previous works may have: when trying to attach meaning, the thinking and intuition is still Newtonian. This is particularly evident in~\cite{Curiel} where physical ideas provide the starting point to get to Lagrangian dynamics. The starting point is that a classical system is one that, roughly, obeys Newton's second law and therefore we have, for a free particle, $\dot{x} = v$ and $\dot{v} = 0$. Unfortunately this is only true in an inertial frame. As~\cite{North} notes, both Hamiltonian and Lagrangian mechanics are coordinate independent: the equations are valid in all reference frames, even non-inertial ones. In fact, that is probably the most notable advantage of the frameworks, that fictitious forces are automatically handled. But none of the Newtonian concepts are coordinate independent: inertial frames, absence of forces and even conservation of energy are all frame dependent. This, again, leads to confusion and we may miss a key element: in the transition from Newtonian to Lagrangian/Hamiltonian mechanics, ~\cite{Curiel} must necessarily expand the reference frames in which the equations are valid (from inertial frames to all frames) at the expense of restricting the class of systems we can describe (e.g. dissipative forces in general require a modification of the Euler-Lagrange equation typically using the Rayleigh dissipation function). It is not clear when and how that transition between models happens. The point is, Newtonian concepts may prevent us from fully understanding Hamiltonian and Lagrangian mechanics.

This attitude of treating mathematical structures in isolation as fundamental objects in a physical theory, instead of tools to capture its formal aspects, is now somewhat pervasive in some areas of physics, but it should be noted that it was not always like that. Even just a century ago, there was a push to find fundamental principles or laws expressed in physical terms that could serve as a starting point for the different theories. Newtonian mechanics, thermodynamics and special relativity serve as good examples. Nowadays, unfortunately, many of the current fundamental theories, including Lagrangian mechanics, Hamiltonian mechanics, quantum mechanics and quantum field theory take as their starting point a particular mathematical structure. That is, we do not know what the value of a Lagrangian means or what the symplectic form represents; we just take them as given. Moreover, a lot of work in theoretical physics nowadays starts by exploring novel abstract mathematical structures and then trying to see whether some physical meaning or prediction can be extracted. We are not going to debate here how and why this shift happened. We merely note that it leaves the semantics of our theories ill defined.\footnote{Which is a nice way to say that we don't really know what we are talking about.} 

It is only natural that, given the circumstances, one would use the only thing that seems to be well defined, the math, and try to draw understanding from there. But here lies the fundamental problem: modern mathematics is not created by physicists to do physics, it is created by mathematicians to do math. The definitions of topology, differentiable manifold, symplectic manifold, Hilbert space and so on are chosen either because they are convenient to prove theorems or because they are convenient to perform calculations.\footnote{In fact, it is common in math to have two such definitions and then prove they are equivalent.} We should not expect them to be a good match for definitions that would be physically meaningful. In other words, there is a very good chance that looking at just the math may lead us, like horses with blinders, in a well-defined direction which is not the physically meaningful one. To be clear, we do not fault the mathematicians as they are not responsible for whether their structures are used appropriately or not within physics.

As all these problems stem from a lack of physical principles that can serve as a foundation for classical Hamiltonian and Lagrangian mechanics, we turn to our recently published work~\cite{AoPPhy1} that aims to identify such starting points.\footnote{We should also mention~\cite{Maudlin} as another attempt to formally derive mathematical structure from physical ideas.} Continuing along the lines of those insights, we fully develop another that is only briefly hinted at in that work: the idea that \textbf{a distribution over a manifold conserves information entropy over time if and only if its elements evolve according to Hamiltonian mechanics.} That is, the single requirement of information entropy conservation not only gives us Hamiltonian dynamics but also the structure of phase space (i.e. a symplectic structure where variables are organized in conjugate pairs).

This insight helps to clearly characterize what Hamiltonian mechanics describes, providing a clear physical meaning to the mathematical structures used. This should be at least of interest to that part of the philosophy community interested in the foundations of classical mechanics. Secondly, information entropy seems to have a connection to the foundations of thermodynamics and statistical mechanics, in the sense that it is used successfully in physics practice, even on systems far from equilibrium (see \cite{Jarzynski, Maes, Horowitz}). Therefore the strong link we establish here between information entropy and Hamiltonian mechanics should be of interest to an even wider audience.

We begin in section 2 by giving a brief summary of statistical distributions in general. These are your typical mathematical tools used to keep track of how charge is distributed in space or how a population of individuals is distributed in age. We briefly introduce information entropy, which measures the number of yes or no questions one has to answer to be able to identify a particular element within a distribution. The main issue is that the density and the information entropy for a distribution depend, in general, on the variables and their units. For example, if we change our units from meters to kilometers a density of $1 kg/m^3$ would change to $10^9 kg/km^3$. In other words, distributions are not in general invariant under change of variables. In section 3 we look at distributions over states. We argue that since states themselves are coordinate independent, the density and information entropy associated with them cannot be coordinate dependent. Moreover, a deterministic and reversible system must be one that preserves information entropy over time (i.e. information about the system at one time is equivalent to information about the system before or after). The requirement of coordinate-independent distributions gives us the symplectic structure (i.e. conjugate pairs of position and momentum) while the requirement of conservation of information entropy gives us Hamiltonian evolution. In section 4 we discuss the result and provide a dictionary that allows us to give precise meaning to the different mathematical symbols, addressing the issues highlighted before. We also show how this understanding allows us to construct a classical analogue of the uncertainty principle.

\section{Distributions}

Our starting point is the standard concept of a statistical distribution. We have a set of elements, also called population, and a set of variables that characterize some properties of those elements. To each element is associated a value for each variable and we want to describe how the population is distributed over the possible values of the variables. For example, people by age and country of residence, amount of material among different compounds, mass or charge over different positions in space. In all these cases, we have a total quantity $M$ that measures how many elements we have. This can be either a discrete quantity, like the number of people we are considering, or a continuous quantity, like the total amount of mass. We also have a set of $n$ variables $\xi^a$ over which the elements are distributed and a function $m(\xi^a)$ that gives us the amount for each possible value such that:
\begin{equation}
\int m(\xi^a) d\xi^1 d\xi^2 ... d\xi^n = M
\end{equation}
The integral is used when some of the variables $\xi^a$ are continuous, and a simple sum would be used over the discrete ones. We can also define the normalized density as:
\begin{equation}
\rho(\xi^a) = \frac{1}{M}m(\xi^a)
\end{equation}
To calculate the amount in a particular region $V$ we have:
\begin{equation}
m_V = \int_V M \rho(\xi^a) d\xi^1 d\xi^2 ... d\xi^n
\end{equation}
The above expression would represent the mass or charge within a particular region, the amount of material of some selected set of compounds or the world population in a particular set of countries and age group. We can also define:
\begin{equation}
\mu(V) =\frac{m_V}{M} = \int_V \rho(\xi^a) d\xi^1 d\xi^2 ... d\xi^n \\
\end{equation}
which gives the fraction of the population within that region.

The normalized distribution $\rho(\xi^a)$, as defined, is a statistical distribution that represents the entirety of the system being described. Yet, we can give it an additional meaning of probability if we run the following thought experiment: suppose we took an element at random, each element with equal chance. That is, we have a uniform distribution on the elements (not the values). What is the probability $P(\xi^a \in V)$ that the values for $\xi^a$ associated with that element are within a particular region $V$? The answer must be $\mu(V)$ and therefore $\rho(\xi^a)$ is the probability density. So $\rho(\xi^a)$ has this double role as a physical statistical distribution and a conceptual probability distribution for our thought experiment. This second role is our link to information theory.

Before proceeding, though, let us give a quick introduction so that we have a clear understanding about what information entropy represents. In the physics community, for example, the concept is often vaguely referred to as ``knowledge" (and sometimes as ``lack of knowledge") which is incorrect and misleading.\footnote{We believe this characterization may be due to Jaynes~\cite{Jaynes} who first introduced the concept of information theory within thermodynamics. Note that we are only going to talk about information entropy and not thermodynamic entropy. We do believe there is a link between the two, but it is not the one Jaynes gives, which we believe to be unphysical. We leave that discussion for later works.} Information entropy was introduced by Shannon~\cite{Shannon} to solve specific engineering problems and therefore, like distance or mass, information entropy is a number that measures a well defined quantity and should be treated as such. What he needed was a way to measure how much information a particular channel is able to transfer and therefore a way to quantify precisely information as number of bits. So the idea is that we have a source that sends messages to a destination, and the message is encoded in zeros and ones. Information entropy is the optimal average number of zeros and ones that must be used to encode the messages. If the source sends fewer bits per message than required by the information entropy, some information will be lost. If the source sends more bits per message than required by the information entropy, some information will be redundant.

\begin{figure}\label{twoQuestions}
	\centering
\begin{tikzpicture}[level distance=1.5cm,
level 1/.style={sibling distance=6cm},
level 2/.style={sibling distance=3cm}, sloped]
\node {Is it red or green?}
child {node {Is it red?}
	child {
		node {red - 11}
		edge from parent node[pos=0.6, align=center, above] {Yes}
	}
	child {
		node {green - 10}
		edge from parent node[pos=0.6, align=center, above] {No}
	}
	edge from parent node[pos=0.6, align=center, above] {Yes}
}
child {node {Is it blue?}
	child {
		node {blue - 01}
		edge from parent node[pos=0.6, align=center, above] {Yes}
	}
	child {
		node {yellow - 00}
		edge from parent node[pos=0.6, align=center, above] {No}
	}
	edge from parent node[pos=0.6, align=center, above] {No}
};
\end{tikzpicture}
\caption{The decision tree for the example in the text. The nodes of the tree represent the questions Bob can pose to Alice to identify the color and the bottom line shows the four different cases with the corresponding encodings.}
\end{figure}
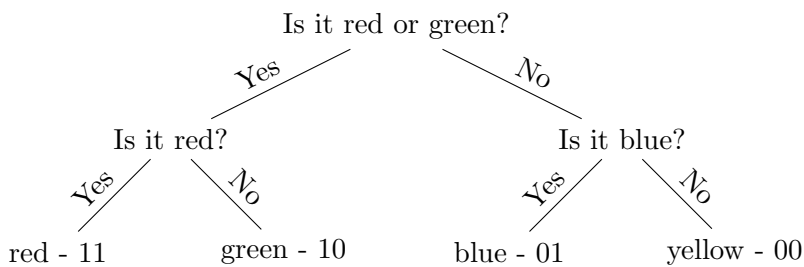

To understand how this works in our context, let us go back to our distribution $\rho(\xi^a)$ and refine our thought experiment. Suppose two individuals, Alice and Bob, have full knowledge of the same distribution. Suppose Alice picks an element at random and Bob needs to know the value of the variable for the element Alice picked. That is, Alice is the source, Bob is the destination and the message is the value of the element Alice picked from the distribution. How long, on average, should the message be? How many ones and zeros? How many yes or no questions must Bob ask Alice before he knows the value? For example, suppose there are 40 balls of which 20 are red and 20 are green. In that case one yes or no question (``Is it green?"), one bit of information, would be sufficient. Suppose there are 40 balls of which 10 red, 10 green, 10 blue and 10 yellow. In this case, two questions would suffice as shown in figure \ref{twoQuestions}, since two yes or no questions, two bits of information, can distinguish 4 cases.

In general, if $I$ is the number of questions, the number of bits, then $C = 2 ^ I$ are the number of cases one can distinguish with those questions. On the other hand, suppose for a particular value $\xi_0$ of a discrete variable we have $\rho(\xi_0)=1/5$, then we know that we'd select $\xi_0$ one in 5 cases. That is, the normalized distribution is the inverse of the number of cases: $\rho(\xi_0) = \frac{1}{C}$.  Therefore the information needed to identify a particular case is $I(\xi^a)=\log \frac{1}{\rho(\xi^a)}$. If we take the expectation value of this expression over the distribution and change the base of the logarithm, we have:
\begin{equation}
I(\rho(\xi^a)) = \int \rho(\xi^a) \ln \left(\frac{1}{\rho(\xi^a)}\right) d\xi^1 d\xi^2 ... d\xi^n =-\int \rho(\xi^a) \ln (\rho(\xi^a)) d\xi^1 d\xi^2 ... d\xi^n
\end{equation}
which is the information entropy of the distribution. This represents the average number of bits one needs to put into a message to tell someone else the value of the element of the distribution that was picked at random.

Technically, given the more convenient choice of the natural logarithm, the unit of information is the nat, which is equal to $\frac{1}{\ln 2}$ bits. However, in the discussion we will refer to it as measured in bits, as we believe that the majority of the readers is more comfortable with the latter unit. Also, technically, for continuous variables the expression has a slightly different meaning: the uniform distribution in a unit volume is assigned zero entropy so $I(\rho(\xi^a))$ represents the relative number of bits to achieve the same level of precision, which can be negative. That is, if we take a real number from a uniform distribution between 0 and $\frac{1}{2}$ we need to lose a bit of information to get to the same precision of a uniform distribution over a unitary interval. But these technical details do not change the general intuition.

The above discussion, while not a formal derivation, should give a sense of what the expression signifies: it measures the \emph{additional} information that would be needed to identify the value associated to an element picked at random from a known population. Distributions that are more spread will have a higher information entropy. What is also important is that it is a precisely defined quantity and that there is only one way to define it. It is the only expression that satisfies three simple requirements (continuity, monotonicity and additivity) that are necessary to measure information, and that is how Shannon originally derived it. Insofar that we want to talk about amounts of information we have only one way to do so.

\subsection*{Densities and change of variables}

There is one feature about densities we need to be fully aware of: in general they are not invariant under change of variables. Intuitively, if we change units the density changes its numerical value. For example, $1$ kg/m becomes $1000$ kg/km. Physically, we use the unit to understand that the overall quantity is the same. But mathematics only keeps track of the number, which changes, and is therefore coordinate dependent. In general, if $\hat{\xi}^a=\hat{\xi}^a(\xi^b)$ is the change of variable, we can recover how the density changes by requiring that the total in each volume is the same no matter what variables are used. That is:
\begin{align*}
\int_V \rho(\xi^b) d\xi^1 ... d\xi^n &= \int_V \rho(\hat{\xi}^a) d\hat{\xi}^1 ... d\hat{\xi}^n \\
&=\int_V\rho(\hat{\xi}^a) \left\|\frac{\partial \hat{\xi}^a}{\partial \xi^b}\right\| d\xi^1 ... d\xi^n
\end{align*}
\begin{equation}\label{density_transformation}
\rho(\xi^b) = \rho(\hat{\xi}^a) \left\|\frac{\partial \hat{\xi}^a}{\partial \xi^b}\right\|
\end{equation}
the density is multiplied by the absolute value of the Jacobian determinant of the transformation. As the Jacobian is in general a function of the variables themselves, this means that whether the density at one point is greater or smaller than at another also depends on the choice of variables.

Similarly, we can see that information entropy is not invariant
\begin{align*}
I(\rho(\xi^b)) &=-\int \rho(\xi^b) \ln (\rho(\xi^b)) d\xi^1 ... d\xi^n \\
&=-\int \rho(\hat{\xi}^a) \left\|\frac{\partial \hat{\xi}^a}{\partial \xi^b}\right\| \ln \left(\rho(\hat{\xi}^a) \left\|\frac{\partial \hat{\xi}^a}{\partial \xi^b}\right\|\right) d\xi^1 ... d\xi^n \\
&=-\int \rho(\hat{\xi}^a) \ln \left(\rho(\hat{\xi}^a) \left\|\frac{\partial \hat{\xi}^a}{\partial \xi^b}\right\|\right) d\hat{\xi}^1 ... d\hat{\xi}^n \\
&=-\int \rho(\hat{\xi}^a) \ln (\rho(\hat{\xi}^a)) d\hat{\xi}^1 ... d\hat{\xi}^n -\int \rho(\hat{\xi}^a) \ln \left\|\frac{\partial \hat{\xi}^a}{\partial \xi^b}\right\| d\hat{\xi}^1 ... d\hat{\xi}^n
\end{align*}
\begin{equation}\label{entropy_transformation}
I(\rho(\xi^b)) =I(\rho(\hat{\xi}^a)) -\int \rho(\hat{\xi}^a) \ln \left\|\frac{\partial \hat{\xi}^a}{\partial \xi^b}\right\| d\hat{\xi}^1 ... d\hat{\xi}^n
\end{equation}
The reason is that we may be changing scale and therefore level of precision required to describe our value. Given that the change in both density and entropy is determined by the Jacobian, we conclude that \textbf{during a change of variable the information entropy of all possible distributions does not change if and only if the density remains the same at every point under the same change of variable}.

To sum up, while we would like to think of a density as a number associated to a point of a space, that is technically incorrect. Given a manifold $\mathcal{M}$ upon which we define the distribution, the latter is \emph{not} $\rho : \mathcal{M} \to \mathbb{R}$ in general. That would be a scalar function, invariant under change of variables, as a point is a point no matter how it is identified. Instead we additionally need a set of functions $\xi^a : \mathcal{M} \to \mathbb{R}$ and only then we can define the density as $\rho : \mathbb{R}^n \to \mathbb{R}$. The density value is defined on the values for a set of variables, not at a point.\footnote{As the density is really a limit, we need to know how that limit is taken, and the coordinates decide that.} That is, the density is undefined without a (differentiable) set of coordinates.

\subsection*{Mathematical tools for distributions}

Different fields in math deal with the coordinate dependency of densities in different ways depending on their aims, which unfortunately leads to a fractured physical understanding given that one needs ideas and results from the different areas. In statistics and information theory one simply accepts the transformation rules and does not try to create invariant objects. This gives us the proper setting to study the statistical properties but it does not help us understand which mathematical elements represent objective physical entities.

In probability theory one starts with three objects: a set of outcomes $\mathcal{M}$, a set of events $\sigma_\mathcal{M}$ where each event is a collection of outcomes, and a probability measure $\mu : \sigma_\mathcal{M} \rightarrow [0,1]$ that assigns a number between zero and one to each event. That is, we do not assign a probability density to the points of $\mathcal{M}$ but we assign a finite probability to finite regions. We can then define a random variable $\xi : \mathcal{M} \rightarrow \mathbb{R}$ as a real-valued function of the outcomes. For each random variable we can define its cumulative distribution function $F_\xi(\xi_0)=\mu(\xi<\xi_0)$ and the probability density function is its Radon-–Nikodym derivative $\rho(\xi) = \frac{dF_\xi}{d\xi}$. That is, the integral is the primal object while the density is derived.

This approach, if applied to statistical distributions, actually makes more physical sense: what we measure is the amount of mass in a finite region and the density is the limit for a smaller and smaller one. The fundamental objects (outcomes, events and probability) are all defined without reference to variables. It also has the advantage that discrete and continuous quantities are treated in the same way. But in probability theory we have no notion of coordinate systems, vectors or any other geometrical objects.

In differential geometry, one starts with a manifold: a set of points $\mathcal{M}$ which can be given local coordinates $\xi^i : \mathcal{M} \to \mathbb{R}$. One defines a tangent space $\mathsf{T}\mathcal{M}$ where vectors live and cotangent space $\mathsf{T}^*\mathcal{M}$ were linear functions of vectors live. Then we define $n$-forms as multi-linear functions of $n$ vectors that return the value associated with the infinitesimal parallelepiped they form. That is $\nu : (\mathsf{T}\mathcal{M})^n \rightarrow \mathbb{R}$. As vectors are coordinate invariant, $n$-forms are coordinate invariant as well so we can write $\mu(V)=\int_V \nu$ with no reference to coordinates. If $e_a$ are the basis vectors associated with $\xi^a$ and $e^b$ are linear functions such that $e^b(e_a)=\delta_a^b$, then we can express each $n$-form as $\rho(\xi^a)e_1\wedge e_2 \wedge ... \wedge e_n$. The density $\rho(\xi^a)$ is the component of the $n$-form expressed in the $\xi^a$ coordinates in a way analogous to components of a vector.\footnote{Technically, the form is linear so it changes with the Jacobian determinant and not its absolute value. But a negative determinant means changing orientation of the region of integration, so fixing that would bring another sign change and $\rho(\xi^a)$ would not change. This is equivalent to saying that some coordinate systems are right handed and some are left handed.}

This approach gives us a way to understand which objects are coordinate independent and which are not. But it has two issues: it does not allow us to easily talk about coordinate dependent operations, such as marginal distributions and information entropy, and it may use definitions that do not map well to physical concepts. For example, a vector is typically defined as a map between a scalar function of the manifold to another scalar function of the manifold, or as an equivalence class of trajectories. It is a stretch to think of fluid velocities or electric fields in those terms.\footnote{For that same reason, we do not use the differential geometry notation $\frac{\partial}{\partial \xi^a}$ and $d\xi^a$ for vector and covector basis because they do not convey the actual physical meaning of these objects as they are used in physics. This is another instance of the math-for-mathematician issue.}

So the situation is that we have a single clear physical object we want to study, a statistical distribution, but different and somewhat disconnected mathematical frameworks to study it. As noted in the introduction, this should not be surprising as the frameworks are defined by mathematicians to solve their own needs. But it is an issue when these are used in physics without careful consideration, as each mathematical framework may only partially capture our physical understanding. And this is precisely why we argue that simply examining the mathematical structure used in a physical theory is not enough to gain a complete picture: mathematics can only check for self-consistency and not for completeness or correctness. What we will need to do, in fact, is use elements from all these frameworks as some insights are better understood in one context and some in another.

\section{Coordinate-independent distributions}

We now turn our attention to what we call an infinitesimally reducible system~\cite{AoPPhy1}. This is a system that can be repeatedly divided into smaller and smaller parts and for which giving a description of the full system is equivalent to giving a full description for all parts. We call particles the limit of this recursive subdivision. Note that particles here are not point-like: they have an infinitesimal size for each dimension along which the system was subdivided. Under this premise each particle is assigned a state from a state space $\mathcal{S}$ and the state of the whole system will be described by a distribution over those particle states. The elements of our distribution are the particles, and the property that characterizes each element is its state.

Additionally, suppose that this system evolves in a way that is deterministic and reversible, that is, the state of a particle at one time is associated with one and only one state at another, future or past, time.\footnote{Note that by reversible we mean the ability to reconstruct the past, the dual of determinism, which we take to mean the ability to predict the future. We do not mean a time symmetric process, where the equation of motion does not change when time is reversed, or a process that can be undone by another.} In this case, all the particles and only the particles that are in the same initial state will be mapped to the same final state: the density at the initial state must be the same as the density at the final state. Moreover, if Alice gave Bob enough information to identify the state of an element at one time, Alice also gave Bob enough information to identify the state of the same element at a future or past time: the information entropy will remain the same over time.

While the physical setting is clear enough, it introduces a puzzle. States will be themselves identified by quantities, which we call state variables. But for states to be physically objective, they must be independent of reference frame. For example, if $s$ is the state, $A$ and $B$ are two reference frames and $\xi^a$ and $\hat{\xi}^b$ are the state variables one would use in the respective reference frames, we must have $s(\xi^a) = s(\hat{\xi}^b)$.  Therefore, since the distribution is defined on the state itself, we must have $\rho(s(\xi^a))=\rho(s(\hat{\xi}^b))$. In this case, it seems we \emph{do} want the density to be defined on the actual space $\mathcal{S}$, and it seems it will have to be invariant under an arbitrary change of state variables. Accordingly, if we want densities and information entropy to be conserved over time, we first need to make sure these quantities are defined in a way that is observer independent, since otherwise they would be conserved in one frame but not in another. But this seems in apparent contradiction with the previous section and we need to clarify that.

For starters, we concede that \textbf{only invariant distributions, those for which the density and information entropy do not change under a change of coordinates, can characterize the state of a system made of infinitesimal parts and its deterministic and reversible evolution.} Since requiring information entropy to be invariant will automatically require the density to be invariant as well and vice-versa, the requirement is one and the same.

Now, if our variables are discrete, all our integrals are sums and all our definitions work: everything is already invariant. On the other hand, if our variables are continuous (i.e. the state space is a manifold) we have a problem as densities and information entropy will not be invariant under an arbitrary change of state variables. Our state space $\mathcal{S}$, somehow, must allow us to write densities that, under coordinate transformations, change with the Jacobian (i.e. something we can integrate on) but also not change at all (i.e. something that is coordinate independent). How can this possibly work?

Suppose $\mathcal{S}$ is a two dimensional manifold. Suppose we have two variables $(q,k)$ such that $k$ uses the inverse units of $q$. For example, if $q$ is meters then $k$ is inverse meters which means an infinitesimal area $dq dk$ will be a pure number. We say $q$ is a coordinate as it is a variable that defines a unit, while $k$ is the corresponding conjugate variable as it uses the inverse unit. Now suppose you change coordinate. For example, $\hat{q}$ will be kilometers and $\hat{k}$ inverse kilometers. We will have $d\hat{q} d\hat{k} = dq dk$. Therefore we will also have $\rho(q,k) = \rho(\hat{q}, \hat{k})$: the density is invariant and so is the information entropy.

And here lies the solution to the puzzle: there is a difference between state variables, quantities used to identify a state, and coordinates, particular state variables that additionally define the system of reference and units. The density and information entropy must be invariant under arbitrary change of coordinates, not arbitrary change of state variables. The insight is that one choice of coordinate can influence more then one state variable. But how many can it influence?

It turns out that if we have a set of variables fixed by a single choice of unit, a single coordinate, then a pair $(q, k)$ is the only possible way. Let $q$ be the coordinate, the variable that defines the unit. Suppose we have $n$ variables $\xi^a$ such that $\xi^1 = q$. The vector $d\xi^a$ representing the infinitesimal displacement has $n$-components. If we change unit $\hat{q}=\hat{q}(q)$ we have two constraints: the unit change and density invariance, which means the Jacobian must be unitary. These must be enough to determine how the vector components change. If not, we would need to specify another relationship between the old $\xi^a$ and the new variables $\hat{\xi}^b$ which means it is not true that the units of all variables are fixed by a single choice of unit for $q$. As the linear system for $d\hat{\xi}^b$ must be solvable, the number of vector components must be equal to the number of constraints and therefore there are exactly two variables $\xi^a = (q, k)$. Then $k$ is the conjungate variable of $q$. There is no other choice.

To sum up, the density and information entropy will be invariant under coordinate transformations but will not be, in general, invariant under all state variable transformations. Physically, the coordinates play a double role. They define the reference frame and also help to identify particle states. It is this double role that gives them a special character because a change in $q$ must induce a change in $k$ as their units and the unit of $\rho$ are not independent. This is the key physical insight that will lead us to Hamiltonian mechanics. Note that symplectic geometry strips away the units and does not elevate densities as primary objects. It is no wonder this insight gets lost. Which goes back to our original point: we need to know what the mathematical structures represent physically and that cannot be deduced from the math itself.

\subsection*{Hamiltonian mechanics for one degree of freedom}

We are now ready to see how conservation of information entropy leads to the phase space of Hamiltonian mechanics. For now, we assume our invariant distribution is characterized by only one coordinate which means two state variables $(q, k)$. While we have discussed units for the state variables, we still need to choose the unit for the density itself. In the same way a normalized density in space is a pure number over a volume (e.g. units of $1/m^3$), a normalized density for particle states will be defined over a region of possible states. By convention, we use $\hbar$ as the unit to describe the number of possible states in a region. We have:
\begin{equation}
m_V = M \mu(V) =\int_V m(q, k) \hbar dq dk = \int_V M \rho(q, p) dq dp
\end{equation}
where we set $p=\hbar k$. We recognize $p$ as conjugate momentum and $k$ the classical analogue of the wave number. Note that, at this point, $\hbar$ is simply a constant that defines the unit for the number of states. It is our choice to set it to a particular value. That is, we do not yet know that there is a ``special number of states'' that represents a ``smallest size'' in some sense. That requires an extra assumption which we will discuss when looking at the classical uncertainty principle.

Now suppose we change variables. In general the Jacobian will be given by:
\begin{equation}
\label{Poisson}
\begin{aligned}
|J| &= \left| \begin{matrix}
\dfrac{\partial \hat{q}}{\partial q} & \dfrac{\partial \hat{q}}{\partial p} \\[2.2ex]
\dfrac{\partial \hat{p}}{\partial q} & \dfrac{\partial \hat{p}}{\partial p} \end{matrix} \right| = \frac{\partial \hat{q}}{\partial q} \frac{\partial \hat{p}}{\partial p} - \frac{\partial \hat{p}}{\partial q} \frac{\partial \hat{q}}{\partial p} &= \{\hat{q}, \hat{p}\}
\end{aligned}
\end{equation}
which we recognize to be the Poisson bracket. If the change of variables is simply a change of coordinates, then the new coordinate $\hat{q}$ is an arbitrary (possibly non-linear) function of just the old coordinate $q$. We have:
\begin{equation}
\label{coordinate_change}
\begin{aligned}
\hat{q} &= \hat{q}(q) \\
\{\hat{q}, \hat{p}\} &= 1 = \frac{\partial \hat{q}}{\partial q} \frac{\partial \hat{p}}{\partial p} \\
\dfrac{\partial \hat{p}}{\partial p} &= \frac{\partial \hat{q}}{\partial q} ^{-1} = \frac{\partial q}{\partial \hat{q}} \\
\hat{p} &= \frac{\partial q}{\partial \hat{q}} p
\end{aligned}
\end{equation}
In general, conjugate momentum $\hat{p}$ will depend on the coordinate $q$ through the function $\frac{\partial q}{\partial \hat{q}}$, which is a constant only in the case the transformation is linear. This is exactly the transformation law for the covariant component of a covector. On the other hand, the density and information entropy are invariant:
\begin{equation}
\label{density_invariance}
\begin{aligned}
\rho(q,p) &= \rho(\hat{q}, \hat{p}) \\
I(\rho(q,p)) &= I(\rho(\hat{q},\hat{p}))
\end{aligned}
\end{equation}
Let us call $(q,p)$ canonical variables if the density is equal to the invariant density. Then two other variables such that $\{\hat{q}, \hat{p}\}=1$ are also canonical since the Jacobian determinant will be unitary.

Note that we have our cake, the density transforms like a Jacobian, and we can eat it too, the Jacobian is one and therefore the density and the information entropy are invariant. This is exactly what we needed and we have also given a precise meaning to all the pieces: the wave number $k$ is the additional state variable that each coordinate must introduce to give invariant densities, $\hbar$ is the unit we use to count the possible states over a region which is the inverse of the unit of the invariant density, canonical variables are those that define the density in the proper unit, a canonical transformation is one that leaves the density invariant and each coordinate transformation is a canonical transformation because conjugate momentum changes like a covector. It is striking how much we were able to get with such a simple starting point of coordinate-independent distributions.

Now it is time to find the equations of motion. We can think of the evolution as a vector field on $\mathcal{S}$ of components $S = (S^q, S^p) = (\frac{dq}{dt}, \frac{dp}{dt})$ that gives the direction in which states move in time. After an infinitesimal time step, we would have:
\begin{equation}
\begin{aligned}
q(t+dt) &= q(t) + \frac{dq}{dt} dt = q(t) + S^q dt \\
p(t+dt) &= p(t) + \frac{dp}{dt} dt = p(t) + S^p dt
\end{aligned}
\end{equation}
The Jacobian for time evolution will be
\begin{equation}
\label{Jacobian_evolution}
\begin{aligned}
|J| &= \left| \begin{matrix}
1 + \dfrac{\partial S^q}{\partial q}dt & \dfrac{\partial S^q}{\partial p} dt \\[2.2ex]
\dfrac{\partial S^p}{\partial q}  dt & 1 + \dfrac{\partial S^p}{\partial p} dt \end{matrix} \right| \\
&= 1 + \left[ \dfrac{\partial S^q}{\partial q} + \dfrac{\partial S^p}{\partial p} \right]dt + O(dt^2)\\
&= 1 + div(S)dt + O(dt^2)\\
\end{aligned}
\end{equation}

If we want the time evolution to transport  densities and to conserve information entropy, the Jacobian must be unitary and therefore $S$ is divergence-free and admits a potential. We have
\begin{equation}
\label{Potential_Hamilton}
\begin{aligned}
S &= \left(\frac{\partial H}{\partial p}, - \frac{\partial H}{\partial q}\right) \\
\frac{dq}{dt} &= \frac{\partial H}{\partial p}  \\
\frac{dp}{dt} &= - \frac{\partial H}{\partial q}  \\
\end{aligned}
\end{equation}
and we recognize Hamilton's equations. Note that the reverse is true as well: any Hamiltonian evolution will transport densities and conserve information entropy due to Liouville's theorem.\footnote{Again, it is not just that the density and the areas are conserved in time: it is that they are the same for all canonical transformations including change of coordinates. That is, we have an objective way to compare them at different points.} Therefore \textbf{Hamiltonian mechanics is precisely the evolution of (classical) distributions over a manifold that preserves information entropy and densities}.

We would like to stress how, once the simple premise of invariant distributions under deterministic and reversible evolution is accepted, all elements of Hamiltonian mechanics are necessary and have a straightforward physical meaning. Not only that: because of the equivalence we know that everything in Hamiltonian mechanics can be understood with just that simple premise. And the premise is not about cotangent bundles of Lie groups of quasi-q spaces on an infinite-dimensional lattice. It is just the simple realization that if we require distributions to have a physically objective reality they must be coordinate independent. The premise is much simpler than the math employed. We can reason intuitively on the premise. We can think about what it means and when it is applicable. We know what we are talking about.

Note how we have not talked about mass, acceleration or forces. None of those Newtonian concepts appeared. And since we were able to get phase space and the equations of motion without them, it means they cannot play any significant role. Note how all the concepts we used, states and parts of a system, are automatically coordinate invariant. Note how it is clear what type of time evolution we are \emph{not} describing: those that are non-deterministic or non-reversible. Those that would concentrate or spread our distributions, those that would change the amount of information to describe a particle state at a set level of precision are ruled out.

Note how each mathematical object found its natural place. Note how we had to cross the boundaries of the different mathematical frameworks to find meaning as the current mathematical frameworks we have do not lend themselves to talk about issues of units and coordinate-invariance of information entropy. Note, though, that our reasoning is logically necessary and therefore, in principle, we should be able to create a physically motivated mathematical framework that would capture these ideas. We just cannot expect mathematicians to magically do it for us.

\subsection*{Distributions over multiple degrees of freedom}

To understand what happens with multiple degrees of freedom, let us first look at change of variables applied to marginal distributions. Suppose we have a distribution $\rho(\xi^a)$ that depends on many coordinates, each with its unit. If the density is coordinate invariant, then it will be so even if we change only one coordinate, say $\xi^1$. As we saw before, then we can find the conjugate variable $\xi^2$ that changes in the opposite way. We can calculate the marginal distribution by integrating over those two quantities.
\begin{equation}
\rho(\xi^b) = \int \rho(\xi^a) d\xi^1 d\xi^2
\end{equation}
where $\xi^b$ are the remaining coordinates. If the variables $\xi^b$ are truly independent from $\xi^1$ and $\xi^2$, then a change of coordinates in the first will not affect the others and vice-versa. This means $\rho(\xi^b)$ must be coordinate invariant. In terms of information theory, invariance of the marginal densities is equivalent to being able to assign a coordinate invariant information entropy to each of the marginal distributions.

We can imagine proceeding iteratively, integrating away one pair at a time. We will find that all our variables are paired $(q^i, p_i)$  with the first ones defining the units and the second ones defining the invariant areas. We call each pair an independent degree of freedom as it constitutes an independent choice of unit. We recognize $\mathcal{Q}$, the manifold charted by $q^i$, as the (poorly named) configuration space. We are going to call it coordinate space as it includes all variables and only the variables that define the units and the reference frame. The state space for the particles, then, is a set of coordinates plus an equal number of conjugate variables that vary like covector components.

The idea, then, is that a distribution is truly coordinate-independent only when the density and information entropy for all the possible marginal distributions on any combination of degrees of freedom are invariant, not just for the joint distribution.

\subsection*{The geometry of phase space}

Now that we have an intuitive sense of the objects we want to describe physically, we can turn to differential geometry to derive the geometrical structure of the state space and then the equations of motion. What we want to capture is that we can integrate densities over two dimensional sub-manifolds in a way that is coordinate invariant. That is, we need to have a two-form $\omega$ such that:
\begin{equation}
\rho_\Sigma = \int_\Sigma \rho(\xi^a) \omega(d\Sigma)
\end{equation}
is coordinate invariant.\footnote{In fact, creating a marginal distribution means giving a foliation (a parametrized family of surfaces that do not intersect and whose union is the whole space) and then calculating the integral as a function of the parameters.}

The requirement that $\omega$ is a two-form, i.e. linear and anti-symmetric, comes from the fact that it needs to represent areas. That is, it needs to be linear because the area of a parallelogram is a linear function of its sides. And if $v^1$ and $v^2$ are vectors and $\omega(v^1, v^2)$ represents the area, then rotating by 90 degrees will give the same area. That is, $\omega(v^1, v^2) = \omega(v^2, -v^1) = -\omega(v^2, v^1)$. Therefore it must be anti-symmetric. It should also be non-degenerate, i.e. there is no direction for which $\omega$ is always zero or we would not be able to define a volume.

We also want the two-form to be closed, that is the integral
\begin{equation}
\int_\Sigma \omega(d\Sigma)
\end{equation}
is zero if $\Sigma$ is a closed surface. To understand why, note that it corresponds to the case where $\rho(\xi^a)=1$. If the surface is a parallelepiped in some variables, each face that gives a positive contribution will have a parallel face that would give an equal and opposite contribution. As the closed integral needs to be zero for all parallelepipeds, it will also be zero for an arbitrary closed surface.

Under a change of variables, the two-form $\omega$ must be invariant. That is, the particle state space $(\mathcal{S}, \omega)$ is a symplectic manifold. And, as any symplectic manifold, we can find Darboux variables $q^i, p_j$ and express the two-form as
\begin{equation}
\label{Symplectic}
\begin{aligned}
\omega &= dq^i \wedge dp_i = d\xi^a\omega_{ab}d\xi^b \\
\omega_{ab} &=  \left[
\begin{array}{cc}
0 & 1 \\
-1 & 0 \\
\end{array}
\right] \otimes I_n =
\left[
\begin{array}{cc}
0 & I_n \\
-I_n & 0 \\
\end{array}
\right]
\end{aligned}
\end{equation}

The components $\omega_{ab}$ let us understand what geometrical features are captured by $\omega$. It corresponds to a two dimensional vector product within each degree of freedom while it corresponds to a scalar product across degrees of freedom. As $\omega$ is preserved during coordinate transformations, the area within each degree of freedom is coordinate independent meaning areas of phase space are invariant and so are the marginal densities and information entropies. Across degrees of freedom, instead, we are preserving the angle meaning two perpendicular degrees of freedom remain perpendicular. But here perpendicularity means independence, so whether two degrees of freedom are independent is now invariant as well. As these concepts are coordinate invariant, they correspond to actual physical objects that all observers can agree on. They are part of the objective physical reality being described.

Now, the fact that we can use symplectic manifolds, or tangent bundles, to track our particle states does \emph{not} mean that conjugate momentum is a one form, in the sense of a map from a vector to a number, defined on the coordinate space. The state space comes as one whole manifold, a set of points identified by $n$ continuous variables $\xi^a$, and the structure it has comes from how units transform, not because the variables identify separate objects. Unfortunately, mathematical structures strip out the units, the knowledge of the double role of coordinates is lost and then a different mathematical concept is used, one that has the same mathematical features of our particle state space but an altogether different meaning and construction. We could, in many cases, use position and velocity as state variables. While these would not be canonical coordinates and thus not a good choice to track our densities, they are still valid state variables and in fact, given that velocity is gauge invariant, very useful ones~\cite{Holten}. But the space has not changed: we still have states. The tangent and cotangent bundle structures, then, are just tracking how conjugate momentum and velocity change differently under coordinate transformation.

\subsection*{Hamiltonian mechanics for multiple degrees of freedom}

As deterministic and reversible evolution will conserve densities and information entropies, it will conserve $\omega$. That is, $\omega(v, w) = \omega(v',w')$ where $v'$ and $w'$ are the vectors evolved in time. In math terms, deterministic and reversible evolution for an invariant distribution is a symplectomorphism. As before, $S = \left(\frac{d\xi^a}{dt}\right)$ is the vector field corresponding to the evolution. The vector components change according to the Jacobian.
\begin{equation}
	v'^a = \partial_b \xi^a(t+dt) v^b = (\delta^a_b + \partial_b S^a dt) v^b
\end{equation}
Therefore we have:
\begin{align*}
v^{a} \omega_{ab} w^{b} &= v'^{a} \omega_{ab} w'^{b}  \\
&= (v^{a} + \partial_{c} S^{a} v^{c} dt) \omega_{ab} ( w^{b} + \partial_{d} S^{b} w^{d} dt) \\
&= v^{a} \omega_{ab} w^{b} + (\partial_{c} S^{a} v^{c} \omega_{ab} w^{b} + v^{a} \omega_{ab} \partial_{d} S^{b} w^{d}) dt \\ &+ O(dt^2)
\end{align*}
If we set $S_{b} \equiv S^{a} \omega_{ab}$, we have:
\begin{equation}
\begin{aligned}
v^{c} w^{b} \partial_{c} S_{b} &- v^{a} w^{d} \partial_{d} S_{a} = 0\\
\partial_{a} S_{b} - \partial_{b} S_{a} &= curl(S_{a}) = 0 \\
S^{b} \omega_{ba} = S_{a} &= \partial_{a}H \\
\frac{dq^i}{dt} &= \frac{\partial H}{\partial p_i}  \\
\frac{dp_i}{dt} &= - \frac{\partial H}{\partial q^i}
\end{aligned}
\end{equation}

We recognize Hamilton's equations for multiple degrees of freedom. Again, we stress how the symplectic structure was recovered by requiring invariant distributions which in turn requires invariant integration over independent degrees of freedom. Note how the requirement was given outside of the symplectic structure itself and therefore it cannot be recovered simply from that.

\subsection*{On the physicality of information entropy}

We want to stress, again, that in general information entropy cannot be taken as an objective physical quantity. It depends on the choice of symbols one uses for encoding: using letters of the alphabet or words in the English language will yield a different information entropy for the same book. It will depend on the choice of ensemble: in game theory one uses credence distributions that depend on the agent and therefore the information entropy will be different for different actors. The nature of the information entropy, then, depends on the nature of the elements, the ensemble and the values over which the distribution is defined.

Moreover, we should note that our construction is not the one typically used in classical statistical mechanics. In that case, the elements are points across $N$ copies of phase space (one set of position and momentum for each particle) and the distribution over them is a probability distribution. In this case one has to be careful and show how the distribution is uniquely defined by objective physical entities. If one interprets said distribution as credence, like \cite{Jaynes}, then one has obvious problems with viewing information entropy as a physical quantity.

Our case, then, is by construction very particular. The ensemble is the state of a physical system. The elements are the physical subdivisions of said system, the particles. The subdivisions are dictated by what parts can be studied independently. The set of values over which the distribution is defined is the state space of the particles. The densities relate what fraction of the whole can be found in a particular particle state. Densities and areas in terms of $q^i$ and $k_i$ are invariant pure numbers, independent of the choice of unit. No ``partitioning" or ``averaging" of phase space is invoked. The information entropy depends on no other external factors: it is an invariant state variable of the ensemble. It is a physical property of the system.

Our statistical distribution does not correspond to the probability distribution of statistical mechanics. Conceptually, it corresponds to a point across the $N$ copies of phase space, a microstate, in the following way: pick a very narrow unitary distribution $\delta(q,p)$; make a copy for each particle so that it is centered around its position and momentum; sum them all.\footnote{Note that if we permute the particles we obtain the same distribution, so this repackaging does not suffer from overcounting.} That is:
\begin{equation}
\rho(q,p) = \sum\limits_{a=1}^N \delta(q - q_a, p - p_a)
\end{equation}
To form a complete link to statistical mechanics we would have to define what the probability distribution is. To form a complete connection to thermodynamics we would have to introduce a link to thermodynamic entropy. These additional steps are outside of the scope of this work. Here we simply want to show how information entropy applied to physical distributions over phase space is indeed a physical quantity that is deeply linked to the notion of deterministic and reversible evolution and to Hamiltonian mechanics.

\section{Discussion}

Hamiltonian mechanics, through Liouville's theorem, has always had an important role in statistical mechanics. What we are showing is that that role is not a coincidence: it is the very defining characteristic. \textbf{Hamiltonian mechanics is a type of statistical mechanics}, one with deterministic and reversible laws.

This approach clarifies what we believe is a source of confusion. One may think that conceptually starting with parts and aggregating them to a whole is equivalent to starting from a whole and dividing it into parts. This is true only in the finite case. In the infinite case we have to start with what is finite (either the whole or the parts) and take the limit (either division or aggregation). We contend that the top-down approach (i.e. we start with a finite whole and keep subdividing) is the correct one to understand classical mechanics.

First because it maps better to what we do in classical physics. We study composite systems (planets, cannonballs, beads on a wire, fluids) which we pretend we can recursively divide. None of those objects has a single position or momentum: they have an (idealized) center of mass with position and momentum given by the expectation values $\int q \rho(q,p) dq dp$ and $\int p \rho(q,p) dq dp$ of the distribution of its parts. A ``point particle'' is really a peaked distribution whose extent can be disregarded for the problem at hand.

Second because it solves a set of conceptual problems: if we start with point-like objects of zero size, how can we get a finite size? Shouldn't we have a discrete topology? How could we have a differentiable structure? We contend that measure, topologies, metrics, differentiable structure and so on are there to remind us that the points are infinitesimal limits, keep track of how they were taken and which aggregations of those limits correspond to a finite whole.

Third because it is conceptually more compatible with the other physical theories. We have seen how elements of special relativity (e.g. coordinate invariance, momentum as covector), statistical mechanics (e.g. joint and marginal distributions), quantum mechanics (e.g. wave number) and classical Hamiltonian mechanics (e.g. phase space, Poisson brackets) already emerged naturally. Furthermore, note that in general relativity point-like particles would be singularities that puncture space-time and would not move along geodesics, while particles as infinitesimal parts would. Note that if the evolution is deterministic and reversible, it means the evolution of the system depends only on itself, which means that the system is isolated. That is, we have an association between isolated system and conservation of energy.\footnote{To be precise, conservation of energy is not an invariant property. The fact that a Hamiltonian exists, though, is an invariant property.} This sounds strikingly close to the first law of thermodynamics.

The larger point, then, is that physics is one and it cannot truly be understood piecemeal. Nature does not separate between dynamical systems and thermodynamics, between topologies and $\sigma$-algebras, between classical systems and quantum systems. The boundaries between all these disciplines are as much unfortunate as they are artificial. There is no real understanding of one without the other. This approach helps to resolve this by giving us a more coherent picture that spans across the different areas, which is not only intellectually more satisfying but it allows us to switch perspective when the intuition from one fails. As a practical example, we can show how understanding the link between Hamiltonian mechanics and information theory leads to a classical analogue for the uncertainty principle.

\subsection*{Classical uncertainty principle}

As we saw, Hamiltonian dynamics not only conserves the energy of individual particles, but it also conserves the information entropy of the distribution as a whole. Intuitively, we can see that there is some link between the spread of a distribution and the number of bits required to identify an element. For example, if we have two uniform distributions with different ranges, then Alice will need to give Bob more information to identify the value from the bigger range. So we ask: what is the distribution that minimizes the spread given a certain amount of information entropy? If $I$ is the set amount of information entropy and $\sigma_q^2 \sigma_p^2 \equiv \int (q-\mu_q)^2 \rho \, dqdp \int (p-\mu_p)^2 \rho \, dqdp$ is the spread, we can use Lagrange multipliers to find out.
\begin{align*}
L = &\int (q-\mu_q)^2 \rho \, dqdp \int (p-\mu_p)^2 \rho \, dqdp \\
&+ \lambda_1(\int \rho dqdp - 1) \\ &+ \lambda_2(- \int \rho \ln \rho \, dqdp - I)\\ 
\delta L = &\int \delta \rho [(q-\mu_q)^2 \sigma_p^2 + \sigma_q^2 (p-\mu_p)^2 + \\ &\lambda_1 - \lambda_2 \ln \rho - \lambda_2 ] dqdp = 0 \\
\lambda_2 \ln \rho = &\lambda_1 - \lambda_2 + (q-\mu_q)^2 \sigma_p^2 + \sigma_q^2 (p-\mu_p)^2 \\
\rho = &e^{\frac{\lambda_1 - \lambda_2}{\lambda_2}}e^{\frac{(q-\mu_q)^2 \sigma_p^2}{\lambda_2}}e^{\frac{\sigma_q^2 (p-\mu_p)^2}{\lambda_2}}\\
\end{align*}
We solve the multipliers and have:
\begin{align*}
\rho = &\frac{1}{ 2 \pi \sigma_q \sigma_p} e^{-\frac{(q-\mu_q)^2}{2\sigma_q^2}} e^{-\frac{(p-\mu_p)^2}{2\sigma_p^2}} \\
I = &\ln (2\pi e\sigma_q\sigma_p)
\end{align*}
The distribution that minimizes the spread, then, is the product of two independent Gaussians. As the entropy is conserved during Hamiltonian evolution the product $\sigma_q^2 \sigma_p^2$ can never be less than the one given by the Gaussian distribution of the same entropy. We have:
\begin{align*}
\sigma_q\sigma_p \geq \exp (I) / 2 \pi e 
\end{align*}
Hamiltonian mechanics is the deterministic and reversible evolution for a distribution and its elements, and therefore it cannot reduce the spread of the distribution indefinitely. If it did, it would start concentrating the density, meaning that multiple states would be mapped to a single state and the dynamics would not be reversible. In this form $I$ is not a fundamental constant but a parameter that depends upon the initial condition. While this is already striking, one can construct a simple argument to turn this into something more~\cite{AoPPhy1}.

Suppose that we have a distribution where the particles move non-deterministically. That is, their motion is partly caused by some other agent in a way that can be characterized by random motion. Now suppose our distribution is at an equilibrium with that random motion. Then the particles change state randomly but the overall distribution remains the same. That is, the random motion simply reshuffles the elements. In this case the minimum achievable for the information entropy of the distribution depends on the random motion itself. The stronger it is, the more spread the distribution has to be, the higher the minimum threshold for the information entropy.

Now suppose that source of chaotic motion is something intrinsic to a physical system, something that cannot be avoided. Maybe it is caused by interaction with the environment, from which the system can never be fully isolated. Or maybe it is caused by some internal motion, which can never be fully accessible. Or maybe it is caused by some background process with space itself, since the vacuum is not just empty space. The point here is not to identify the source of this non-deterministic motion, the point is to show that assuming one exists does not sound unreasonable.

If one buys into the argument, one can posit that there exists an $I_0$ associated with this chaotic process that acts as the theoretical limit for the information entropy of our distributions. The farther the information entropy of our distribution is from that value, the more we can assume that the parts of the distribution evolve deterministically. The closer it is, the more our assumption breaks down.

We can then decide to measure volumes in phase space in terms of this minimum spread, which means setting the value of $\hbar$ (or $h = 2 \pi \hbar$) according to the following relationship:
\begin{align*}
\frac{\hbar}{2} &= \frac{h}{4\pi} \equiv  \frac{e^{I_0}}{2 e \pi} \leq \sigma_q\sigma_p
\end{align*}
 We refer to this as the classical uncertainty principle~\cite{AoPPhy1}.

\subsection*{Point particles vs infinitesimal parts}

At this point one may still argue that point particles and infinitesimal parts are both abstract concepts and both factually incorrect (i.e. planets are neither points nor made of infinitesimal parts), so why should we prefer one concept over the other?

Proper understanding of a physical theory means answering, among others, the question: why does the theory work experimentally? Why does the prediction match our data? And the answer cannot merely be: because it does. The problem is not just that the answer begs the question; it is that no known theory works experimentally for all possible systems in all possible settings. Newtonian gravitation does not predict gravitational lensing; general relativity is not expected to hold at the smallest scales. There are no ``right" theories in an absolute sense, only ones that agree with experiments in a set of circumstances. So it is not clear why in some cases disagreement with experiment would discard the theory altogether while in others it would not.

We need to rephrase the question as: why does the theory work experimentally in these cases, but not those others? In other words, what is the realm of applicability of our theory? If the theory disagrees with experimental data within the realm of expected applicability, then the theory will be either discarded or amended in its applicability. In practice, physicists and engineers learn by examples which particular theory is suitable for what system, without a clear-cut line of demarcation.\footnote{To illustrate this with an anecdote: it is not uncommon to find individuals, even physics professors, who will insist that Hamiltonian mechanics is perfectly equivalent to Newtonian mechanics. When asked to provide the Hamiltonian for a damped harmonic oscillator, they will right away note that they cannot because it is a dissipative system. Then they will typically either admit that the two are different, or that they are still the same but only when they are both applicable.} Ideally, it would be preferable to find a set of necessary conditions that the system needs to satisfy for the theory to be applicable. This would give a clear line of demarcation within which empirical adequacy is both expected and justified. The description of the system in the theory matches experimental verification because, in those circumstances, the system can be considered to satisfy those conditions. If we chose to apply the theory where those conditions were not satisfied, it would not be the theory's fault that it gave incorrect predictions. It would be our fault for incorrectly applying it.

To us, this is not just an abstract intellectual question. It is a practical matter. I have a system to study and have a choice of different models (e.g. Hamiltonian mechanics, Newtonian mechanics, quantum mechanics, Markov processes, ...). How do I know which one to pick?  What should I tell a student?

If we characterize classical Hamiltonian mechanics as describing point-like objects, whose state is described by position and momentum, and that move according to a specified set of equations, it is not clear why it would apply to a planet and to a speck of dust in a vacuum, but not to a speck of dust in a fluid or to an electron. The size does not seem to matter while the circumstances seem to. Saying that Hamiltonian mechanics describes point-particles does not clarify what physical objects can be described by the theory: it does not help us understand what classical systems are. Moreover, why should a point-particle be constrained to follow those equations? A damped harmonic oscillator does not.\footnote{We should again stress that, because every first-order system of equations can be turned into a Hamiltonian one\cite{AllSystemsAreHam}, we can always find a conservative system that in a suitable non-inertial frame ``looks like'' (i.e.~it has the same trajectories of) a non-conservative one. For example, if we see a body coming to a full stop, it could be because drag was acting on it, because it ejected part of its mass, or simply because we accelerated to a comoving frame. The kinematic variables (i.e.~position, velocity and acceleration) are not enough to define whether the system is actually non-conservative, especially if we do not know what reference system we are in. Therefore an equation in terms of $x$, $\dot{x}$ and $\ddot{x}$ is not enough to know we have ``a system under drag''. It is the dynamic variables (i.e.~momentum, energy, force and mass) that provide that characterization: it is the loss of energy through heat that tells us there was drag. Therefore, if one looks closely at Hamiltonian realizations of equations that come from non-conservative systems, such as \cite{chandrasekar2007lagrangian}, one will find non-standard relationships between position, velocity, momentum and energy. At least one of these relationships will have to break down at the equilibrium (e.g.~diverging momentum or energy). There is no way around this because Hamiltonian systems cannot have attractors, while a true dissipative system must have one.} Why is Hamiltonian mechanics not applicable there? Therefore this characterization does not help us identify the realm of applicability.

If we instead characterize classical Hamiltonian mechanics as the deterministic and reversible motion of objects reducible to infinitesimal parts, we understand when the model choice is appropriate. Can we pretend that the system we are studying, under the conditions we are studying it, is made of infinitesimal parts that can be studied independently? In the case of an electron the answer is no. In the case of a planet or a speck of dust, if we are studying only the overall motion, then yes. Can we pretend that the evolution is deterministic and reversible? In the case of the speck of dust in a fluid, no. For a planet and speck of dust in a vacuum, yes. With this characterization, then, a classical system is one that can be considered to be infinitesimally reducible. It is furthermore Hamiltonian if the evolution is deterministic and reversible.\footnote{In the same vein, we find that it is Newtonian if it is purely kinematic (i.e. spatial motion identifies the state), it is Lagrangian if it is both Hamiltonian and Newtonian and so on~\cite{AoPPhy1}.} The structure of phase space and the equations of motion are a direct consequence of those conditions. It is clear what the realm of applicability of the theory is and its empirical validity is justified within that realm.

Therefore we prefer the characterization presented here as it gives a set of necessary conditions from which we can rederive the theory. These act as a clear demarcation for when the theory is expected to match experimental data (i.e.~its realm of applicability), which we believe is a necessary element for a full understanding of any physical theory.

\subsection*{A dictionary}

We mentioned several times how this approach allows us to give precise physical meaning to the different mathematical objects. We leave in table \ref{dictionary} a summary of these connections in the form of a math-to-physics dictionary.
\begin{table}[h!]
	\centering
	\begin{tabular}{c p{0.3\textwidth} p{0.5\textwidth} }
		& Name & Meaning\\ 
		\hline 
		& Classical particle & an infinitesimal part of the system (i.e. the limit of recursive division) \\ 
		$\mathcal{S} =\mathsf{T}^*\mathcal{Q}$ & Phase space \newline (Cotangent bundle) & the set of all possible states for particles upon which invariant distributions can be defined \\
		$\xi^a$ & State variables \newline (Unified coordinates) & the set of variables needed to identify the state of a particle \\ 
		$\rho(\xi^a)$ & Density & the amount of material for a given particle state (i.e. in the limit of the recursive division)\\ 
		$I(\rho(\xi^a))$ & Information entropy & the number of bits to identify an element of the distribution to a unitary level of precision\\ 
		$q^i$ & Coordinate & a state variable that also defines a unit \\
		$\mathcal{Q}$ & Coordinate space \newline (Configuration space) & the space charted by all coordinates \\
		$k_i$ & Conjugate coordinate & a state variable that uses the inverse unit of the corresponding coordinate \\
		$\hbar$ & & the unit for quantifying the number of possible states within a single degree of freedom \\
		$p_i=\hbar k_i$ & Conjugate momentum & a state variable that together with the corresponding coordinate defines invariant ranges of possible states \\
		$(q^i, p_i)$ & Canonical variables \newline (Canonical coordinates) & a set of variables for which the density is invariant and expressed over units of $\hbar$ per degree of freedom\\ 
		& Canonical transformation & a change in state variables that does not change the value of the density\\ 
	\end{tabular}
	\caption{Dictionary between mathematical and physical objects.}
	\label{dictionary}
\end{table}

We can also better understand what physical requirement each mathematical structure is used for. Manifolds are used to represent a set of physical objects that can be identified by $n$ continuous variables.  A differentiable structure on a manifold is used to define densities. Non-differentiable transformations, in fact, do not define a Jacobian and therefore cannot express the density in the new variables. A symplectic manifold is used to define coordinate invariant densities. Non-canonical transformations, in fact, have non-unitary Jacobians and therefore the density would change its value.

\section{Conclusion}

In this paper we have shown how classical Hamiltonian mechanics coincides with conservation of information entropy. The structure of phase space is required to be able to define densities and information entropy in a way that is coordinate invariant, and therefore physically meaningful. The equations of motion and the existence of the Hamiltonian are required to conserve densities and information entropy over time. This clarifies the link between Hamiltonian mechanics, statistical mechanics and deterministic and reversible motion.

In the narrower context, it allows us to understand what physical object is represented by each mathematical one. In the broader context, it allows us to make deeper connections among different areas of math and physics, and pushes us to rethink our starting points. In particular, it tells us that classical particles should not be considered point-like objects (upon which entropy and densities cannot be defined), but the limit of recursive subdivisions (i.e. infinitesimal cells of phase space).

This approach may point to a reorganization of fundamental physics on a more physically rigorous conceptual footing and more meaningful mathematical grounds which we believe is not only possible, but long overdue.

\section*{Acknowledgements}

We would like to thank Josh Hunt, David J. Baker and Gordon Belot for feedback. We would especially like to thank Mario Hubert: without his inspiration and guidance this work would have not been possible. This article is part of a larger project, Assumptions of Physics, that aims to identify a handful of physical principles from which the basic laws can be rigorously derived.\footnote{See http://assumptionsofphysics.org/book for further details}

\bibliographystyle{alpha}

\bibliography{bibliography}{}

\newcommand{\etalchar}[1]{$^{#1}$}
\begin{thebibliography}{GLNMSO{\etalchar{+}}12}

\bibitem[Bar14]{Barrett1}
Thomas~William Barrett.
\newblock On the structure of classical mechanics.
\newblock {\em The British Journal for the Philosophy of Science},
  66(4):801--–828, 2014.

\bibitem[Bar18]{Barrett2}
Thomas~William Barrett.
\newblock Equivalent and inequivalent formulations of classical mechanics.
\newblock {\em The British Journal for the Philosophy of Science}, 2018.

\bibitem[CABB18]{AoPPhy1}
Gabriele Carcassi, Christine~A. Aidala, David~J. Baker, and Lydia Bieri.
\newblock From physical assumptions to classical and quantum {H}amiltonian and
  {L}agrangian particle mechanics.
\newblock {\em Journal of Physics Communications}, 2(4):045026, 2018.

\bibitem[Cof14]{Coffey}
Kevin Coffey.
\newblock Theoretical equivalence as interpretative equivalence.
\newblock {\em The British Journal for the Philosophy of Science}, 65:821--844,
  11 2014.

\bibitem[CSL07]{chandrasekar2007lagrangian}
VK~Chandrasekar, M~Senthilvelan, and M~Lakshmanan.
\newblock On the {L}agrangian and {H}amiltonian description of the damped
  linear harmonic oscillator.
\newblock {\em Journal of Mathematical Physics}, 48(3):032701, 2007.

\bibitem[Cur14]{Curiel}
Erik Curiel.
\newblock Classical mechanics is {L}agrangian; it is not {H}amiltonian.
\newblock {\em The British Journal for the Philosophy of Science},
  65(2):269--–321, 2014.

\bibitem[GLNMSO{\etalchar{+}}12]{AllHamFreeParticle}
E.~Galindo-Linares, E.~Navarro-Morale, G.~Silva-Ortigoza, R.~Suárez-Xique,
  M.~Marciano-Melchor, R.~Silva-Ortigoza, and E.~Román-Hernández.
\newblock Any {H}amiltonian system is locally equivalent to a free particle.
\newblock {\em World Journal of Mechanics}, 2(5):246--252, 2012.

\bibitem[Jar17]{Jarzynski}
Christopher Jarzynski.
\newblock Stochastic and macroscopic thermodynamics of strongly coupled
  systems.
\newblock {\em Phys. Rev. X}, 7:011008, Jan 2017.

\bibitem[Jay57]{Jaynes}
Edwin~T Jaynes.
\newblock Information theory and statistical mechanics.
\newblock {\em Physical review}, 106(4):620, 1957.

\bibitem[LN88]{AllSystemsAreHam}
C.~A. Lucey and E.~T. Newman.
\newblock On the construction of {H}amiltonians.
\newblock {\em Journal of Mathematical Physics}, 29(11):2430--2433, 1988.

\bibitem[Mau14]{Maudlin}
Tim Maudlin.
\newblock {\em New Foundations for Physical Geometry: The Theory of Linear
  Structures.}
\newblock Oxford University Press, 2014.

\bibitem[MN03]{Maes}
Christian Maes and Karel Neto{\v{c}}n{\`y}.
\newblock Time-reversal and entropy.
\newblock {\em Journal of statistical physics}, 110(1-2):269--310, 2003.

\bibitem[Nor09]{North}
Jill North.
\newblock The ``structure'' of physics: A case study.
\newblock {\em The Journal of Philosophy}, 106(2):57--88, 2009.

\bibitem[PHS15]{Horowitz}
Juan~MR Parrondo, Jordan~M Horowitz, and Takahiro Sagawa.
\newblock Thermodynamics of information.
\newblock {\em Nature physics}, 11(2):131, 2015.

\bibitem[Sha48]{Shannon}
Claude~Elwood Shannon.
\newblock A mathematical theory of communication.
\newblock {\em The Bell System Technical Journal}, 27(3):379--423, 7 1948.

\bibitem[vH07]{Holten}
J.~W. van Holten.
\newblock Covariant {H}amiltonian dynamics.
\newblock {\em Phys. Rev. D}, 75:025027, Jan 2007.

\end{thebibliography}

\end{document}